# Behind the Counter: Exploring the Motivations and Barriers of Online Counterspeech Writing

Kaike Ping*

Computer Science, Virginia Tech, Blacksburg, Virginia, United States, pkk@vt.edu

Anisha Kumar*

Computer Science, Virginia Tech, Blacksburg, Virginia, United States, anishak@vt.edu

Xiaohan Ding

Computer Science, Virginia Tech, Blacksburg, Virginia, United States, xiaohan@vt.edu

Eugenia Rho#

Computer Science, Virginia Tech, Blacksburg, Virginia, United States, eugenia@vt.edu

Current research mainly explores the attributes and impact of online counterspeech, leaving a gap in understanding of who engages in online counterspeech or what motivates or deters users from participating. To investigate this, we surveyed 458 English-speaking U.S. participants, analyzing key motivations and barriers underlying online counterspeech engagement. We presented each participant with three hate speech examples from a set of 900, spanning race, gender, religion, sexual orientation, and disability, and requested counterspeech responses. Subsequent questions assessed their satisfaction, perceived difficulty, and the effectiveness of their counterspeech. Our findings show that having been a target of online hate is a key driver of frequent online counterspeech engagement. People differ in their motivations and barriers towards engaging in online counterspeech across different demographic groups. Younger individuals, women, those with higher education levels, and regular witnesses to online hate are more reluctant to engage in online counterspeech due to concerns around public exposure, retaliation, and third-party harassment. Varying motivation and barriers in counterspeech engagement also shape how individuals view their own self-authored counterspeech and the difficulty experienced writing it. Additionally, our work explores people's willingness to use AI technologies like ChatGPT for counterspeech writing. Through this work we introduce a multi-item scale for understanding counterspeech motivation and barriers and a more nuanced understanding of the factors shaping online counterspeech engagement.

CCS CONCEPTS • **Human-centered computing** • **Collaborative and social computing** • **Empirical studies in collaborative and social computing**

**Additional Keywords and Phrases:** Behavior Change, Social Media/Online Communities, Empirical study that tells us about people, Method, Qualitative Methods, Quantitative Methods, Survey

## 1 INTRODUCTION

In today's digital age, social media platforms serve as key spaces for public discourse [37, 47, 117, 160]. While these platforms enable swift dissemination of ideas, they also serve as cultivators for hate speech [25, 101], cyberbullying [5], and harassment [24, 119]. The effectiveness of mitigating online hate through moderation by human moderators and automated systems can vary [57, 61]. Deletion or banning users can sometimes disperse rather than dispel hateful speech [25], or potentially conflict with First Amendment rights in the United States [62]. For example, the practice of deplatforming users for sharing hateful views can simply push offenders to less regulated online spaces [72]. While tech companies continue to combat online hate through traditional moderation methods, limitations in these approaches have led scholars to examine the potential for user-driven counterspeech [16, 42, 120, 133].

---

* These authors contributed equally to this work.
# Corresponding author.

Counterspeech is defined as direct responses to derogatory or harmful content, intended to undermine or refute hateful messages [127, 130]. Since 2016, tech companies like Meta (Facebook), Google (YouTube), and Twitter have partnered with NGOs around the world to foster counterspeech initiatives against online hate [86, 134]. Such focus on user-driven counterspeech efforts highlights the significance of individual and community roles in regulating online spaces [136]. For instance, the transformation of Megan Phelps-Roper, a former member of the extremist Westboro Baptist Church, stands as a testament to the profound impact that counterspeech can have [26]. Phelps-Roper was a 23-year-old legal assistant who regularly posted on Twitter on behalf of the Westboro Baptist Church, which is widely considered as a hate group [156]. In response to her hateful tweets against Jews, David Abitbol, a 50-year-old Jerusalem-based web developer decided to directly engage with her on the platform. Instead of mirroring her hostility or mocking her, Abitbol responded with humor, empathy, and questions, aiming to humanize those she vilified. His constructive counter engagement not only challenged Phelps-Roper's antisemitic views, but also led to a complete reversal of her stance [107]. What started out as a mere tweet in response to a message rooted in hatred gradually undermined and dismantled Phelps-Roper's convictions. This specific case affirms the powerful role that counterspeech can play to enact positive change against online hate. However, successfully engaging in online counterspeech can pose various challenges for individuals. People may vary in their individual motivations and barriers that affect their decision to engage in counterspeech in the first place. Understanding the factors that drive users like Abitbol, and the barriers preventing broader participation in online counterspeech, remains a significant research gap [20, 45, 99].

### 1.1 Motivation of Research Questions

To date, prior research has primarily focused on understanding the content of online counterspeech [43, 53, 104, 134] and its impact on the broader social media ecosystem [68]. There is a dearth of research on what motivates or deters users from participating in it. Our work fills this gap by examining how underlying motivations and barriers influence how often people engage in online counterspeech (**RQ1**).

Demographic factors, such as age [79, 124, 155], gender [79, 124, 151], and race [61, 63] are associated with how people interact online. In the context of counterspeech, research has shown that a counter-speaker's race [106] can influence how others perceive the effectiveness of the counter-speaker's attempt to counter a hateful post [106, 137]. While such studies are a promising start, there remains a lack of knowledge on how broader demographic variables shape people's willingness to engage in online counterspeech. Our work addresses this gap by comprehensively examining how demographic factors affect the likelihood of engaging in online counterspeech across a wide variety of topics **(RQ2)**.

Furthermore, how people feel about their counterspeech [17], the difficulty experienced when writing it [70], and their perception of its effectiveness [74], can influence their willingness to respond to online hate. For example, users' **satisfaction** with their online counterspeech increases when they feel supported by other users in their efforts to challenge hateful actors [17]. For others, the process of writing counterspeech can be daunting even with community support. Simply, the sheer **difficulty** in crafting a counter-message can potentially deter users' willingness to respond to a perpetrator [70]. Similarly, the perceived **effectiveness** of one's counterspeech may affect user's willingness to engage in online counterspeech. The belief that one's words have the power to stop or lessen the offenders' harmful actions can motivate a proactive stance against online hate [74]. However, the perceived effectiveness of one's counterspeech can vary greatly among individuals, with a multitude of factors influencing their belief in its impact. A myriad of variables may shape why people may be *satisfied* with their counterspeech, why they find it *difficult* to write one, and whether they believe it to be *effective* in mitigating online hate. In this work, we examine these factors, specifically, how motivations and barriers influence users' experience in responding to hateful posts, namely their perceived **satisfaction** with their counterspeech, the level of **difficulty** they experience in writing it, and their perception of its **effectiveness** (**RQ3**).

Social media platforms are increasingly adopting artificial intelligence (AI) technologies, like large language models (LLMs), to foster more respectful online interactions [118]. Platforms such as Nextdoor and Quora are now using AI to prompt users to revise potentially provocative or policy-violating posts. For instance, Nextdoor has introduced an OpenAI-powered feature that suggests edits to users' posts to prevent inflammatory language [1, 109].

Although these AI tools aim to cultivate safer online discussions, it remains unclear how users view these interventions and their impact on their interaction with other users. Moreover, users' pre-existing attitudes towards counterspeech and their reasons for engaging or avoiding confrontation with hateful actors online may affect their readiness to embrace AI assistance. By examining the factors that motivate or discourage users from using AI in counterspeech-writing, we can gain deeper understanding into AI's potential to support users in addressing online hate. Our study examines this relationship between individuals' motivations to engage in counterspeech and their openness to using AI tools for such purposes (**RQ4**). In summary, we ask the following research questions:

RQ1. What motivations and barriers influence how often people engage in online counterspeech?
RQ2. How do demographic variables shape people's motivations and barriers in online counterspeech engagement?
RQ3. How do people's motivations and barriers in online counterspeech engagement influence:
   a. how satisfied they are with their counterspeech?
   b. how difficult they find it to write counterspeech?
   c. how they perceive the effectiveness of their counterspeech?
RQ4. How are people's motivations and barriers in online counterspeech engagement associated with their willingness to use AI assistance?

To answer these questions, we conducted a pre-registered survey (N = 458) across English-speaking participants in the United States. Our survey examines key motivations and barriers that underlie why people do or do not engage in online counterspeech, how often they write counterspeech on social media, and their willingness to use AI to help them write counterspeech. In addition, in our survey we showed participants three randomly selected hate posts from a pool of 900 online hate speech posts across five topics: race, gender, religion, sexual orientation, and disability. Participants were asked to respond to each of the three hate posts with a counterspeech. We then asked follow-up questions to understand their perceptions and experience of writing counterspeech (satisfaction, difficulty, and perceived effectiveness of one's counterspeech) in response to the hateful posts.

## 1.2 Overview of Research Findings and Contributions

Our findings show a significant relationship between exposure to online hate and the likelihood of engaging in online counterspeech. Individuals who have personally encountered online hate are often prompted to use counterspeech as a means to directly confront offenders or as an emotional outlet. In contrast, those less frequently exposed to online hate are more likely to engage in counterspeech to signal inclusion to others (RQ1). Demographic factors and social media experiences significantly shape one's motivations and barriers to engage in counterspeech (RQ2). For instance, younger individuals, women, those with higher education levels, and regular witnesses to online hate report more concerns about public exposure, retaliation from the perpetrator, and additional harassment. Notably, even with these concerns, such individuals still find their self-authored counterspeech more effective and satisfying (RQ3).

This is in stark contrast to those who feel more emotionally burdened when writing counterspeech or those who question their ability to write it effectively; these respondents often find their self-authored counterspeech less satisfying, more challenging to write, and less effective. This sense of ineffectiveness is further linked to a higher likelihood of turning to AI for assistance in crafting counterspeech (RQ4). Our results differentiate between users with versus without prior experience using AI tools like ChatGPT. Users with prior experience using AI tools are more inclined to use AI assistance in writing counterspeech to signal inclusivity, but less likely to do so for self-defense. On the other hand, those new to AI tools are more open to using AI for counterspeech writing, especially if they fear retaliation from perpetrators, but this concern does not translate into using AI-mediated counterspeech writing to defend those who are close to them.

We contribute to HCI research by offering a comprehensive understanding of the various factors that shape why people do or do not engage in online counterspeech. We developed and validated a multi-item measure for motivations and barriers for engaging in online counterspeech, demonstrating its significant influence on both the experience of writing counterspeech and peoples' perceptions towards their self-authored counterspeech. These validated measures allow a structured approach for researchers to examine online counterspeech dynamics for future studies. Second, we extended the scholarly discourse on the demographic and experiential factors that motivate or deter people from engaging in online counterspeech. While prior scholarship has mostly focused on counterspeech strategy and content [43, 53, 104, 134], our work contributes to the knowledge of how social, demographic, and personal experiences impact people's counterspeech writing experiences as well as their motivations and barriers to engage in it. By offering insights into how the motivations and barriers for engaging in counterspeech differ among various social groups, our findings can inform the development of counterspeech tools that are tailored to the needs of diverse users [103]. Finally, we contribute to the understanding of what influences people's openness to use AI assistance for writing online counterspeech. This contribution is particularly timely as the role of AI in moderating online communities becomes increasingly prominent. Such insights can help tech companies and researchers to better envision the potential applications and limitations of AI in assisting users countering online hate.

## 2 RELATED WORK

### 2.1 The Role of Counterspeech in Mitigating Online Hate Speech

Counterspeech operates through various mechanisms in mitigating online hate [22]. It can act as a social sanction, increasing the social cost of those who disseminate hate speech and thus discouraging its spread [43, 53, 134]. It can also counter harmful narratives by presenting alternative viewpoints [30, 104]. The efficacy of counterspeech is widely debated [8, 104, 134]. Schieb and Preuss (2016) supports the effectiveness of counterspeech, demonstrating that it can lead to the deletion of hateful posts and even elicit apologies from hateful actors [134]. Their findings suggest that this effectiveness is amplified when counter-speakers outnumber those spreading hate speech, especially if the online community holds moderate views [134]. By contrast, Miškolci et al. (2018) raises questions about the direct impact of counterspeech, that it does not necessarily deter hateful actors from posting hateful content [104]. However, scholars note that a single counterspeech authored by one user often gains visibility among a wider online audience, thereby serving as a catalyst that inspires onlookers to initiate their own counter-responses [104, 113]. The rhetorical style and tone of counterspeech matters too. For example, counterspeech that adopts an empathetic tone has been shown to be particularly effective in leading to offenders deleting their racist and xenophobic tweets [62]. In addition to tone [62, 106, 125, 132], other counterspeech strategies include fact-sharing [17, 112], open denunciation [132, 146], and posing counter-questions [125, 132, 141].

While these studies contribute to the knowledge of counterspeech characteristics and their effectiveness, they often overlook the social and demographic backgrounds of those who use it. Few studies have examined how a counter-speaker's race and online presence might influence their impact of counterspeech [106, 137]. For instance, Munger et al. experimented on Twitter using bots designed to appear as either black or white individuals with varying levels of online status, as indicated by their follower counts [106]. Their findings revealed that counterspeech from a high-status white male bot led to a significant reduction in the use of racist slurs by the original hate speech authors. Despite these initial insights, little is known about the motivations and barriers that influence why or how often people engage in online counterspeech. Our work seeks to fill this gap.

## 2.2 Understanding User Motivations and Barriers in Online Counterspeech Engagement

The success of counterspeech as a remedy to hate speech lies in individuals' readiness to act [17, 18]. Yet, current research falls short in examining *why* people decide to engage in counterspeech or opt to stay bystanders. In comparison, bystander motivation and behavior in cyberbullying are well-researched [147], offering valuable insights that inform this current study.

Researchers highlight strong parallels between bystander reactions to cyberbullying and those faced by people encountering online hate [90, 113, 129]. Hate speech attacks individuals on the basis of social identifiers such as race, gender, and sexual orientation [65, 81, 135]. This differs[1] from cyberbullying, which is characterized by derogating or threatening individuals without necessarily disparaging their social identity [152]. Despite these differences, the challenges and barriers to countering online hate are similar to those in cyberbullying contexts [42]—both involve online users witnessing harmful or hateful behavior [15] and deciding whether to intervene [39]. Likewise, reasons that might deter an individual from defending a bullied peer—fear of exposure, retaliation, or the emotional toll—are similar to the hesitations one might feel when confronting online hate speech [114]. Hence, to develop a comprehensive and nuanced understanding of what drives or dissuades people from engaging in online counterspeech, our study synthesizes insights from prior research in bystander motivation in cyberbullying. Drawing on this body of work, we develop asset of survey variables to delve into the motivations and barriers potentially influencing online counterspeech engagement in the following section.

### 2.2.1 *Motivations for Engaging in Online Counterspeech.*

**M1. Supporting Kin**: Studies show that bystanders are more proactive in countering cyberbullying when they have close emotional or social ties with the victim [15, 39], with a similar trend seen in those countering online hate due to strong connections with friends and family [33].

**M2. Supporting Others**: The motivation to support others in general, as opposed to specific groups or individuals, can be traced back to theories of social responsibility and collective efficacy [129]. Collective efficacy refers to the belief that one's actions can contribute to the greater good, influencing community outcomes [58]. Studies have shown that people are more likely to engage in prosocial behavior when they perceive a moral obligation toward a broader community [36]. In a study examining prosocial online behavior, researchers demonstrate that collective efficacy drives individuals to engage more frequently in altruistic activities [161]. Similarly, those who report feeling close to an online community are more likely to defend someone being targeted by online harassment [33, 34]. Such findings imply that individuals may engage in counterspeech not just to protect themselves or their kin, but for others as well.

---

[1] While a single incident of online hate speech can result in repeated victimization of targets as utterances can have widespread reach on digital platforms, cyberbullying is generally defined to require sustained, long-term exposure on the victim [143, 149].

**M3. Supporting Self**: Engaging in counterspeech can be a deeply personal act [126], especially when individuals feel directly targeted or harmed. However, motivations for self-defense are often nuanced. Guo and Johnson's study shows that users often underestimate the impact of online hate on themselves compared to its impact on others [60]. This perception could potentially influence users' motivation to engage in counterspeech for self-defense, as they may unknowingly downplay the harm directed towards them. Research also shows that personal experiences of online harm or targeted attacks also influence individuals' decisions to counter online hate [144]. Considering these complexities, we include "Supporting Self" as a variable for understanding motivations for counterspeech as it aims to capture the reason that might influence individuals to standup for themselves against online hate speech.

**M4. Confronting Hate**: The urgency to confront hateful or harmful behavior plays a critical role in motivating bystanders to intervene, both in the contexts of online hate speech [54] and cyberbullying [15]. For instance, bystanders are more likely to intervene when the bullying behavior is perceived as more severe [4, 41]. Similarly, the likelihood of bystanders challenging online harassment directly correlates with how menacing they perceive the harassment to be [90]. Hence, we include motivation to confront hateful behavior or people as a variable for engaging in online counterspeech.

**M5. Educating Ignorance**: Researchers have identified ignorance as one of the many factors contributing to the spread of online hate speech, as lack of awareness can lead people to adopt a narrow-minded view of others in society [27]. Hence, various non-profit and educational organizations [16] have advocated education as a strategy to counter online hate over banning users or online censorship [30, 150]. In line with this, Buerger et al. found that counter-speakers are often motivated to educate perpetrators of online hate as to why their message is harmful [18].

**M6. Signaling Inclusion**: Willingness to engage in counterspeech can often be influenced by a desire to signal inclusion, particularly within online communities [54]. Empathy emerges as a key factor in this context: research shows that individuals with higher levels of empathy are not only more inclined to stand up against online hate speech to protect the victim [129, 147], but also to signal a sense of inclusion and community cohesion [66, 95, 147].

**M7. Issue Focus**: Research shows that bystanders are more likely to intervene when the subject matter directly concerns social groups or issues that are important to them [113]. For instance, studies have found that bystanders were more willing to confront misogynist hate speech as compared to homophobic hate speech [113]. This suggests that motivation to engage in counterspeech may depend on issues or topics users are particularly passionate about.

**M8. Venting Emotions**: Studies show that exposure to online incivility often trigger emotion-focused coping strategies, such as venting [98, 128]. Emotional responses, such as anger or frustration, may provoke individuals to "blow off steam" by countering the hate speech they encounter. Carlo et al. further supports this notion, indicating that emotional instability positively correlates with the adoption of emotion-focused coping strategies, which can manifest as aggressive counter responses [19]. Given these findings, we consider "Venting Emotions" as a potential motivation variable for engaging in online counterspeech.

**Motivation Variables (M1-M8)**: In the survey, we presented the motivation variables to participants as statements M1-M8 as shown in Table 1. Participants were asked to indicate the extent to which each factor motivated them to write counterspeech on social media (*How much do the following factors motivate you to write a counterspeech on social media?*) with response options ranging from 1 (None at all) to 5 (A great deal).

Table 1: Motivation Variables and Questionnaire Items

| No | Motivation Variables | Questionnaire Items |
|---|---|---|
| M1 | Supporting Kin | When I feel the need to stand up for people I care about (e.g., family, close friends) |
| M2 | Supporting Others | When I feel the need to stand up for people in general |
| M3 | Supporting Self | When I feel the need to stand up for myself |
| M4 | Confronting Hate | When I want to confront a hateful person or behavior |
| M5 | Educating Ignorance | When I want to educate an ignorant person |
| M6 | Signaling Inclusion | To signal that I stand for inclusion |
| M7 | Issue Focus | When it concerns issues or topics I care about |
| M8 | Venting Emotions | When I want to blow off steam |

*2.2.2 Barriers to Writing Counterspeech on Social Media.*

**B1. Fear of Public Exposure**: Fear of public exposure can play a crucial role in bystander inaction, particularly when intervening would mean revealing oneself to a larger online audience [15]. Studies on online harassment have shown that the larger the audience size, the less inclined bystanders are to take action [15, 64, 90, 96, 111]. Similarly, the public nature of online platforms may deter individuals from engaging in online counterspeech due to fear of public exposure.

**B2. Fear of Perpetrator Retaliation**: Similarly, fear of retaliation from a harasser can significantly influence bystander motivations [10]. A study by Balakrishnan in 2018 found that 40% of bystanders chose not to intervene in instances of cyberbullying due to fears of retaliation [9].

**B3. Fear of Third-Party Harassment**: In addition to retaliation from the perpetrator, users also frequently express concerns about harassment from third parties [76, 129], as confronting online hate can also influence the likelihood of becoming a target of hate speech from others [33]. This phenomenon is supported by Ernst et al.'s 2017 study, which revealed that counterspeech in YouTube comments often attracted additional hateful remarks from third parties [49].

**B4. Time Concern**: Perceived time investment can deter users from engaging in online conflicts [12, 48]. This is reflected in commonly expressed views like "arguing on Facebook is a waste of time" [139]. Hence, it is plausible that concerns about time commitment could discourage users from engaging in online counterspeech, even if they are otherwise inclined to do so.

**B5. Emotional Burden**: While positive emotions like empathy have been found to increase bystander intervention in cyberbullying [52, 110], negative emotional burden could deter such actions [138]. Buerger's work on online activists highlights the emotional toll that counter-speaking can exert, especially when users voluntarily undertake these activities [17]. As a result, for some individuals the emotional cost of online counterspeech can outweigh the perceived benefits, leading to inaction.

**B6. Skill Gap**: According to social cognitive theory, self-efficacy plays a crucial role in bystander decisions [4, 41, 157]. In cyberbullying, bystanders are more likely to intervene when they feel capable and have the necessary resources to help [29]. However, bystanders are less likely to act when they think that other bystanders are more competent than themselves [85]. Likewise, users who feel less equipped to write an effective counterspeech might feel more reluctant to do so.

**B7. Engagement Unqualified**: Freis et al. and Rudnicki et al., highlight that bystanders may refrain from intervening if they feel that it is not their place to interject [52, 129]. Bystanders are also less likely to act in ambiguous situations [85]. Similarly, Piliavin et al.'s work shows that bystanders who witness only the aftermath of a harassment are less likely to intervene than those who see the entire situation unfold, potentially due to feeling less qualified to intervene due to a lack of contextual awareness [121].

**B8. Engagement Reluctance**: A general reluctance to engage in social media discourse can extend to counterspeech, with some individuals more hesitant to enter challenging or confrontational online conversations [38, 124].

**B9. Engagement Ineffective**: Wong et al.'s work shows that the perceived effectiveness of an intervention significantly influences bystanders' willingness to intervene in cases of online harassment [4, 41, 157]. While online counterspeech has been demonstrated to offer support to victims and encourage further counter-responses [43, 53, 134], there remains skepticism about its ability to genuinely alter the perpetrator's attitudes or behaviors [8, 104, 134]. Hence, the perception that one's counterspeech may be ineffective could deter one from engaging in such activities.

**Barrier Variables (B1-B9)**: Similar to the motivation variables, we presented the barrier variables to survey participants as statements B1-B9 (Table 2), and asked them to answer the following question: "*How much do the following factors deter you from writing a counterspeech on social media?*" - with response options ranging from 1 (None at all) to 5 (A great deal).

Table 2: Barrier Variables and Questionnaire Items

| No | Barrier Variables | Questionnaire Items |
| --- | --- | --- |
| B1 | Fear of Public Exposure | I fear being publicly exposed |
| B2 | Fear of Perpetrator Retaliation | I'm afraid of retaliation from the perpetrator |
| B3 | Fear of Third-Party Harassment | I'm afraid that I will be harassed by people (other than the perpetrator) |
| B4 | Time Concern | I don't want to spend time on this |
| B5 | Emotional Burden | Writing a counterspeech is emotionally burdensome |
| B6 | Skill Gap | I don't know how to write an effective counterspeech |
| B7 | Engagement Unqualified | I feel that it's not my place to engage in counterspeech |
| B8 | Engagement Reluctance | I don't like to engage in social media conversations |
| B9 | Engagement Ineffective | I feel that my counterspeech would not make a difference |

## 2.3 The Role of Artificial Intelligence (AI) in Online Counterspeech Engagement

The role of AI, particularly LLMs, in counterspeech research has traditionally focused on detecting hateful speech [28, 54, 68], and generating [88, 122, 131, 148], or evaluating [40, 51, 78, 162] counterspeech. LLMs excel at processing vast amounts of data quickly, alleviating the emotional toll on moderators and users who would otherwise have to detect, respond to, or report hateful online content manually [32, 118]. However, these models are nonetheless limited in their capacity to discern subtle nuances [13, 108] or cultural contexts [75] in hate speech, or distinguish between implicit and explicit forms of hate speech [84, 119]. These constraints often lead to detection failures, resulting in false positives or negatives [61], thus calling the need for better human-AI collaboration in mitigating online hate [83, 84].

More recently, researchers have focused on using LLMs to generate human-like counter-responses to hate speech, using various metrics for measuring the quality of AI-generated counterspeech, such as informativeness [28], politeness [131], and grammatical diversity [162]. However, the complex nature of hate speech, including subjective perceptions of hate [100] highlights the need for human-AI collaboration, not only in detecting hate speech, but also responding to it [33]. Currently, there is a notable lack of research on how people perceive the role of AI assistance with online counterspeech writing. Understanding how users feel about their own counterspeech or what aspects of writing a counterspeech are difficult for them, can better inform the design of AI tools for such purposes. Our research fills this gap by exploring how people's motivations and barriers influence their willingness to use AI assistance when writing online counterspeech. By doing so, our work aims to contribute to a more comprehensive understanding of AI's potential role in online counterspeech engagement, and by large, in combating online hate.

## 3 METHODS

We conducted a pre-registered survey[2] (N = 458) across English-speaking participants in the U.S. to examine key motivations and barriers that underlie why people do or do not engage in online counterspeech, how often they write counterspeech on social media, and their willingness to use AI to help them write counterspeech. The survey showed participants three different examples of hate speech randomly selected from a topically diverse pool of 900 hateful posts and asked them to respond by writing counterspeech in response to each of the three hate posts.

### 3.1 Selecting Hateful Posts

To curate a balanced and representative sample of hate speech for our survey, we sourced hateful posts from three prominent online hate datasets: the ETHOS dataset [105], the Multi-Target Counter Narrative Dataset [50], and the Multilingual and Multi-Aspect Hate Speech Analysis (MLMA) collection [116]. We randomly selected hateful posts across five commonly occurring topics from the combined corpus: gender, religion, disability, sexual orientation, and race. To avoid over or under representation of a specific topic, we balanced our dataset by manually examining all instances of hate speech posts to ensure topical relevance. This resulted in a total of 900 hateful posts across the five topics: race (183), gender (183), religion (182), sexual orientation (182), and disability (170).

### 3.2 Survey Design and Variables

The survey was designed using Qualtrics and consisted of (a) a consent form (b) relevant background information about hateful speech and counterspeech, (c) three hateful posts and questions pertaining to them, (d) questions about past online hate speech experience, frequency of writing counterspeech online, and motivations as well as barriers to writing online counterspeech, (e) questions about prior use of ChatGPT, perceived usefulness of ChatGPT, as well as willingness of using such AI tools to aid in counterspeech writing, and finally (f) demographic and social media use questions. We illustrate the survey flow in Figure 1. All survey questions are presented in the appendix.

The consent form informed participants that they were being invited to a study to evaluate the efficacy of counterspeech to hateful posts on social media, as well as informing them of the potential psychological risks due to the offensive nature of hateful speech. Then, participants were provided with definitions of hateful speech, counterspeech, as well as examples of effective counterspeech.

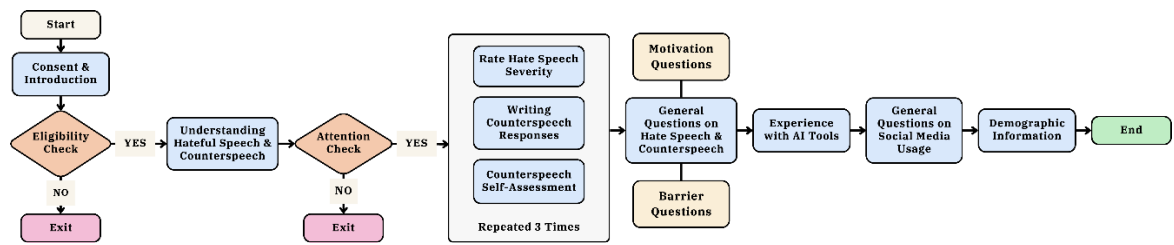

Figure 1: Flowchart of the Survey Process

Following this, participants were shown three unique hateful posts randomly selected from the set of 900 hate posts described in 3.1. For each hateful post, participants were prompted with "*Imagine you are a user of an online group on*

---

[2] The survey was preregistered on Open Science Framework (OSF): https://osf.io/rzmg3/?view_only=6b2fd3a3d42b4b25a37f014612fac18a

*social media. Another user (perpetrator) in the group posted the following. Do you consider this post to be hateful?"* If they answered *Yes*, participants were also asked to rate the hatefulness of each post using a four-point scale, with the question, *"How hateful do you find this post?"* Response options ranged from (1) A little to (4) A great deal.

Participants were then prompted to respond to the hateful post shown. The survey asked, "Please write a counterspeech to this post. The goal is to further reduce hateful behavior from the perpetrator." Participants were then asked to rate their **satisfaction**, perceived **difficulty**, and perceived **effectiveness** of each counterspeech they wrote using a five-point Likert scale. While prior research does not provide specific measures for these variables, given their significance in relation to our study, we developed corresponding response measures. Specifically, we asked, *"How satisfied are you with the counterspeech that you've written?"* (satisfaction), *"How difficult was it to write this counterspeech?"*, and *"How effective do you think your counterspeech would be in preventing the perpetrator from engaging in further hateful behavior?"* with response options ranging from 1- 5 (extremely dissatisfied-extremely satisfied; extremely difficulty-extremely easy; not effective at all-extremely effective).

Finally, participants answered questions related to motivations and barriers to writing online counterspeech, frequency of writing online counterspeech, and willingness to use ChatGPT to write counterspeech on social media. We did not ask users to use AI in writing their counterspeech in the survey. However, our study serves as an initial examination of whether users are willing to use AI for counterspeech, and the factors that shape this willingness. Table 3 lists all variables included in our survey. We asked participants' opinions in the rest of the survey using the conventional 5-point Likert scale, a standard in social science research [35], except for binary response questions (such as prior usage of ChatGPT) and demographic inquiries.

Table 3: Survey Variables

| Independent Variables | Control Variables | Dependent Variables |
|---|---|---|
| **Barriers** | **C1. Demographics** | **RQ1** |
| B1: Fear of public exposure | Age | Frequency of writing online counterspeech |
| B2: Fear of perpetrator retaliation | Gender | |
| B3: Fear of third-party harassment | Ethnicity | **RQ2** |
| B4: Time concern | Education level | Satisfaction |
| B5: Emotional burden | Sexual orientation | Difficulty |
| B6: Skill gap | Political View | Effectiveness |
| B7: Engagement unqualified | **C2. Social Media Behavior & Experience** | **RQ3** |
| B8: Engagement reluctance | Social media commenting frequency | Willingness to use ChatGPT to write counterspeech |
| B9: Engagement ineffective | Use of real name on social media | |
| **Motivations** | Prior experience of online hate speech target | |
| M1: Supporting kin | Frequency of encountering online hate speech | |
| M2: Supporting public | **C3. Prior Use & Perception of ChatGPT** | |
| M3: Supporting self | Prior use of ChatGPT | |
| M4: Confronting hate | Perceived usefulness of ChatGPT | |
| M5: Educating ignorance | | |
| M6: Signaling inclusion | | |
| M7: Issue focus | | |
| M8: Venting emotions | | |

## 3.3 Recruitment

Participants were recruited via Prolific, limited to U.S.-based, English-speaking adults with approval ratings above 95%. All participants were warned about potentially harmful content in the survey. Of the initial 536 respondents, we excluded those who failed attention checks or failed to complete the survey, resulting in a final sample of 458 participants. The average survey completion time was 15 minutes with a compensation rate of $12/hour.

### 3.4 Analysis

**RQ1 What motivations and barriers influence how often people engage in online counterspeech?**

To address RQ1, we performed a linear regression analysis to examine the factors that influence peoples' frequency of writing counterspeech on social media. The dependent variable was participants' self-reported frequency of writing counterspeech, which was measured on a five-point Likert scale in response to the question "*How often do you write counterspeech online?*" The independent variables were the 9 barrier and 8 motivation variables, prior experience being a target of online hate speech, as well as control variables relating to social media behavior and experience as well as demographics. To detect multicollinearity, we also calculated the variance inflation factor (VIF) for each independent variable, with a VIF value greater than 5 indicating a serious multicollinearity problem [115].

**RQ2 How do demographic variables shape people's motivations and barriers in online counterspeech engagement?**

To answer RQ2, we conducted a structural equation model (SEM) to investigate the effects of key demographic factors and social media behavior on peoples' motivations and barriers in engaging in online counterspeech [82]. We first performed an exploratory factor analysis (EFA) using principal axis factoring with oblimin rotation to group the items related to the barriers and motivators into latent variables [87]. We then conducted a confirmatory factor analysis (CFA) using maximum likelihood estimation to test the validity of the model derived from the EFA [87]. Finally, we performed the SEM using maximum likelihood estimation to estimate the path coefficients and test the hypotheses. We assessed the model fit using chi-square, comparative fit index (CFI), root mean square error of approximation (RMSEA), and standardized root mean square residual (SRMR) [73]. We report the standardized coefficients, standard errors, p-values, and R-squared for each endogenous variable in the model.

**RQ3 How do people's motivations and barriers in online counterspeech engagement influence: (a) how satisfied they are with their counterspeech, (b) how difficult they find it to write counterspeech, and (c) how they perceive the effectiveness of their counterspeech?**

In our study, the SEM was adapted to include both two-item and one-item variables. Following Bollen's (1989) guidelines [14] and Daniel's (2021) recommendations [97], our model employed Diagonally Weighted Least Squares (DWLS) estimators, ensuring methodological soundness for two-item variables. Additionally, following the approach by Hayduk et al. (2012) [67], we used a one-item variable, "time", to capture its unique influence on engagement in counterspeech. The two-item and one-item approach was methodological, aiming to account for the complexity of counterspeech engagement. Using two-item and one-item variables for latent constructs is not uncommon, as Wright [158, 159], Blalock [80], Duncan [44], and Heise [69] have done so. This approach has the benefit of allowing more latent variables to be modeled, enhancing theoretical sophistication, and statistical control [67].

To address RQ3, we used the same SEM approach as in RQ2, but included peoples' perceived satisfaction, difficulty, and effectiveness towards their self-written counterspeech as key dependent variables. We examined how these dependent variables were related to the barrier and motivation variables. We followed the same approach for model assessment and reporting as in RQ2. To ensure a comprehensive analysis for RQ3, we incorporated one- and two-item latent variables such as 'time,' to discern their unique influence on engagement with counterspeech, despite their lack of association with other variables. This approach is in line with the work of several scholars who have effectively used one to two-item variables for latent constructs, thereby enriching theoretical depth and statistical control in their studies [44, 69, 80, 158, 159]. To ensure methodological rigor in our approach, we employed Diagonally Weighted Least Squares (DWLS) estimators, as recommended by [14, 97], and conducted robustness checks for our one-item latent variable following the guidelines of [67].

**RQ4 How are people's motivations and barriers in online counterspeech engagement associated with their willingness to use AI assistance?**

To address RQ4, we conducted a linear regression model to examine the relationship between peoples' barriers and motivations for engaging in counterspeech on social media and their willingness to use AI technology, such as ChatGPT, for this purpose. Similar to RQ1, the independent variables consisted of the nine barriers and eight motivation items, while demographic factors as well as social media use and experience variables were used as control variables. The dependent variable was captured via a five-point Likert scale in response to the question, "*If you were writing counterspeech on social media, would you use artificial intelligence technology like ChatGPT to assist you?*" Given that some participants had no prior use of AI technologies like ChatGPT, we conducted two subgroup analyses: one for participants with prior experience using ChatGPT (N=296) and another for those without (N=162). For those who have used ChatGPT before, an additional control variable for perceived usefulness was included.

**Qualitative analysis of open responses**: Finally, to answer RQ4 with more depth, we analyzed the participants' open-ended responses to the following, "*Why would or wouldn't you use artificial intelligence technology like ChatGPT to help you write a counterspeech?*" Using an inductive open-coding approach [145] we first coded the open responses, allowing codes and themes to emerge from the data. We then conducted axial coding to organize and refine the codes and themes to understand how they connected to each other [145]. We then used memoing to make sense of the emerging codes and connections between codes and themes. Throughout this process, three authors discussed emerging themes and connections between codes and themes, and used Cohen's kappa to evaluate the inter-rater agreement for the codes across the authors [153].

## 4 RESULTS

A total of 458 participants (50.6% female, mean age: 40.3±13.3) completed the survey. More demographic details are in Table 1, Appendix. On average, the length of counter speech authored by the participants was 41.6 words.

RQ1 results demonstrate that people who have been targets of online hate speech (targets) tend to engage in counterspeech on social media more frequently than those who do not have such experience (non-targets). This higher engagement stems from a desire to emotionally vent and to confront hateful persons or behaviors. On the other hand, for those who have never been a target of online hate speech, signaling inclusion is one of the main drivers for engaging in online counterspeech.

In RQ2, we show that motivations and barriers influencing counterspeech engagement can be grouped into five latent variables - Fear, Time Concern, Emotional & Skill Barrier, Engagement Hesitation, and Motivation. People's demographic backgrounds and social media experiences impact their willingness to engage in online counterspeech across these five latent variables. Specifically, women, highly educated users, and those frequently encountering online hate are less likely to engage due to fear and emotional/skill barriers, while older, more liberal users, and those who have been targeted by online hate exhibit greater overall motivation for counterspeech participation.

RQ3 results show that motivations and barriers associated with counterspeech engagement significantly impact people's writing experience and perception of their own counterspeech. Individuals more apprehensive about writing counterspeech due to retaliation, public exposure, and third-party harassment were more likely to perceive their own writing as effective and satisfying. Conversely, people with greater emotional and skill-related barriers in writing counterspeech were less likely to perceive their counterspeech as effective and also found it more difficult to write counterspeech to the hateful posts shown in the survey. Surprisingly, individuals who scored higher on the Engagement Hesitation latent factor were significantly more likely to perceive their own counterspeech as effective.

Finally, RQ4 results demonstrate that prior experience of using AI tools like ChatGPT significantly influences people's willingness to use such tools to write online counterspeech. People who do versus do not have experience using AI tools also differ in terms of specific motivations and barriers that underly why they would or would not use AI for counterspeech writing. Prior users are more willing to use AI to signal inclusion to others through counterspeech, but not to defend themselves. In contrast, non-users (those who have never used ChatGPT or similar tools) are more willing to use AI to help them write counterspeech when they fear retaliation from the perpetrator, but they are less inclined to rely on AI to defend friends and family. We delve more deeply into these differences in our open response analyses in section 4.4.

### 4.1 What motivations and barriers influence how often people engage in online counterspeech (RQ1)?

Our linear regression results (Table 4) show four key factors significantly associated with how frequently individuals engage in counterspeech on social media (* $p < .05$; ** $p < .01$; *** $p < .001$). Notably, having been a target of online hate speech emerged as the strongest predictor that drives people to frequently engage in online counter speech ($b = .353$, $\beta = .170$, $p < .001$). Individuals who have been a victim of online hate speech in the past engaged in counterspeech on social media significantly more often than those without such experience. Likewise, those who have higher motivation to engage in online counterspeech in order to stand up for others (other than kin) engaged in online counterspeech more frequently as well ($b = .104$, $p = .033$). Similarly, a stronger motivation to confront hateful persons or behaviors was also significantly related to increased counterspeech activity ($b = .123$, $p = .007$).

On the other hand, reluctance to engage in social media conversations was the strongest negative predictor of how often people engaged in counterspeech on social media ($b = -.172$, $p < .001$), suggesting that peoples' aversion towards engaging in social media conversations in general trump their desire to intervene. We also controlled for how often participants comment on social media in general, which was positively associated with the frequency of writing counterspeech on social media platforms ($b = .146$, $p < .001$). Demographic factors and other motivation and barrier variables were not statistically significant.

Table 4: Linear Regression Results for Counterspeech Writing Frequency on Social Media (N=458)

| | | *B* | β | Std. Error | t value | VIF |
|---|---|---|---|---|---|---|
| Motivations | M1: Supporting kin | 0.030 | 0.038 | 0.043 | 0.689 | 2.481 |
| | **M2: Supporting others** | **0.104** | **0.134** | **0.049** | **2.134*** | **3.291** |
| | M3: Supporting self | -0.025 | -0.033 | 0.038 | -0.655 | 2.085 |
| | **M4: Confronting hate** | **0.123** | **0.160** | **0.046** | **2.692**** | **2.948** |
| | M5: Educating ignorance | 0.021 | 0.029 | 0.043 | 0.493 | 2.868 |
| | M6: Signaling inclusion | 0.039 | 0.054 | 0.037 | 1.048 | 2.216 |
| | M7: Issue focus | 0.002 | 0.002 | 0.047 | 0.033 | 2.782 |
| | M8: Venting emotions | 0.062 | 0.073 | 0.034 | 1.799 | 1.381 |
| Barriers | B1: Fear of public exposure | 0.029 | 0.037 | 0.038 | 0.764 | 2.012 |
| | B2: Fear of perpetrator retaliation | -0.018 | -0.023 | 0.045 | -0.404 | 2.781 |
| | B3: Fear of third-party harassment | -0.008 | -0.011 | 0.042 | -0.194 | 2.750 |
| | B4: Time Concern | -0.046 | -0.064 | 0.032 | -1.417 | 1.703 |
| | B5: Emotional burden | 0.046 | 0.059 | 0.035 | 1.291 | 1.720 |
| | B6: Skill gap | -0.040 | -0.048 | 0.035 | -1.147 | 1.472 |
| | B7: Engagement unqualified | -0.042 | -0.052 | 0.038 | -1.120 | 1.802 |
| | **B8: Engagement reluctance** | **-0.172** | **-0.241** | **0.032** | **-5.400**** | **1.665** |
| | B9: Engagement ineffective | -0.011 | -0.014 | 0.031 | -0.339 | 1.507 |
| SNS | **Past experience of online hate speech target** | **0.353** | **0.170** | **0.077** | **4.586**** | **1.148** |
| | Frequency of encountering hateful content | 0.074 | 0.072 | 0.039 | 1.914 | 1.166 |
| | **Social media commenting frequency** | **0.146** | **0.144** | **0.040** | **3.611**** | **1.320** |
| | Use of real name on social media | 0.037 | 0.062 | 0.022 | 1.690 | 1.133 |
| Demographic | Age | 0.000 | 0.004 | 0.003 | 0.097 | 1.209 |
| | Gender | -0.019 | -0.009 | 0.077 | -0.241 | 1.203 |
| | Ethnicity | -0.092 | -0.045 | 0.076 | -1.216 | 1.154 |
| | Education level | 0.042 | 0.030 | 0.053 | 0.796 | 1.166 |
| | Sexual orientation | -0.081 | -0.033 | 0.097 | -0.835 | 1.289 |
| | Political views | 0.052 | 0.060 | 0.034 | 1.507 | 1.304 |
| | (Intercept) | 1.078 | / | 0.324 | 3.331** | / |

Adjusted R-squared = 0.4532; $F(27, 430) = 15.03$, $p < .001$

**Variability in Counterspeech Engagement Among Targets and Non-Targets of Online Hate Speech**. Given that people who have been targets of online hate speech (targets) in the past tend to engage in counterspeech on social media more frequently than those without such experience (non-targets), we conducted a subgroup analysis to understand how counterspeech motivations and barriers differed between the two groups. We conducted two regression models: non-target group, N=276, and target group, N=182. Results are shown in Table 5 (* p < .05; ** p < .01; *** p < .001). All VIF values were below 3.70.

Targets of Online Hate Speech: Targets engage in counterspeech more often if they have stronger desires to confront hateful persons or behavior (b = .165, p = .045), or to emotionally vent (b = .123, p = .029). However, the greater the reluctance to spend time to engage in online counterspeech (b = -.115, p = .039) and the stronger the perception that counterspeech is ineffective in general (b = -.125, p = .022), the less often do targets engage in counterspeech.

Non-Targets of Online Hate Speech: By contrast, non-targets engage in counterspeech more often when they have a strong desire to signal inclusion to others through counterspeech (b = .116, p = .020). Non-targets also engage in more counterspeech if they frequently encounter online hate (b = .105, p = .031), and comment more frequently on social media in general (b = .176, p < .001). These two variables are insignificant for targets. Furthermore, non-targets are less likely to participate in counterspeech if they generally prefer to avoid social media conversations (b = -.199, p < .001).

This variable was not significant for targets.

Table 5: Factors Affecting Counterspeech Writing Frequency in Targets and Non-Targets of Online Hate Speech

|  |  | Non-target group (n=276) | | | Target group (n=182) | | |
|---|---|---|---|---|---|---|---|
|  |  | *B* | β | t value | *B* | β | t value |
| Motivations | M1: Supporting kin | 0.047 | 0.069 | 0.906 | 0.058 | 0.065 | 0.786 |
|  | M2: Supporting others | 0.094 | 0.135 | 1.514 | 0.115 | 0.135 | 1.383 |
|  | M3: Supporting self | -0.012 | -0.017 | -0.247 | -0.013 | -0.016 | -0.205 |
|  | **M4: Confronting hate** | 0.057 | 0.084 | 1.064 | **0.165** | **0.194** | **2.025*** |
|  | M5: Educating ignorance | 0.040 | 0.061 | 0.783 | -0.045 | -0.057 | -0.598 |
|  | **M6: Signaling inclusion** | **0.116** | **0.172** | **2.338*** | -0.062 | -0.086 | -1.044 |
|  | M7: Issue focus | -0.057 | -0.077 | -0.997 | 0.124 | 0.138 | 1.470 |
|  | **M8: Venting emotions** | 0.046 | 0.057 | 1.031 | **0.123** | **0.149** | **2.202*** |
| Barriers | B1: Fear of public exposure | 0.009 | 0.014 | 0.198 | 0.062 | 0.075 | 0.999 |
|  | B2: Fear of perpetrator retaliation | -0.056 | -0.080 | -0.951 | 0.012 | 0.014 | 0.166 |
|  | B3: Fear of third-party harassment | 0.027 | 0.041 | 0.507 | -0.037 | -0.046 | -0.503 |
|  | **B4: Time Concern** | -0.021 | -0.033 | -0.521 | **-0.115** | **-0.151** | **-2.086*** |
|  | B5: Emotional burden | 0.022 | 0.031 | 0.509 | 0.060 | 0.073 | 0.914 |
|  | B6: Skill gap | 0.005 | 0.006 | 0.105 | -0.076 | -0.090 | -1.282 |
|  | B7: Engagement unqualified | -0.035 | -0.047 | -0.734 | -0.016 | -0.020 | -0.260 |
|  | **B8: Engagement reluctance** | **-0.199** | **-0.316** | **-4.927*** | -0.096 | -0.123 | -1.740 |
|  | **B9: Engagement ineffective** | 0.072 | 0.108 | 1.894 | **-0.125** | **-0.160** | **-2.309*** |
| SNS | **Frequency of encountering online hate speech** | **0.105** | **0.110** | **2.169*** | 0.033 | 0.030 | 0.490 |
|  | **Social media commenting frequency** | **0.176** | **0.190** | **3.532*** | 0.086 | 0.082 | 1.264 |
|  | Use of real name on social media | 0.043 | 0.081 | 1.596 | 0.038 | 0.061 | 1.024 |
| Demographic | Age | -0.003 | -0.048 | -0.948 | 0.008 | 0.085 | 1.240 |
|  | Gender | -0.075 | -0.041 | -0.792 | 0.152 | 0.070 | 1.096 |
|  | Ethnicity | -0.167 | -0.092 | -1.826 | 0.008 | 0.004 | 0.059 |
|  | Education level | -0.027 | -0.022 | -0.403 | 0.039 | 0.025 | 0.395 |
|  | Sexual orientation | -0.048 | -0.019 | -0.358 | -0.081 | -0.035 | -0.555 |
|  | **Political views** | -0.021 | -0.027 | -0.495 | **0.178** | **0.187** | **2.980**** |
|  | (Intercept) | 1.431 | / | 3.760*** | 0.442 | / | 0.717 |
| Adjusted R-squared |  | 0.409 | | | 0.469 | | |
| F-test |  | F(26, 249) = 8.307, p < .001 | | | F(26, 155) = 7.151, p < .001 | | |

### 4.2 How do demographic variables shape people's motivations and barriers in online counterspeech engagement (RQ2)?

To answer RQ2, we conducted a structural equation model (SEM) to investigate the effects of key demographic factors and social media behavior on people's motivations and barriers in engaging in online counterspeech. Prior to constructing our model, we first examined the underlying structure of the motivations and barriers influencing counterspeech engagement using exploratory factor analysis (EFA) and confirmatory factor analysis (CFA) [82]. These analyses helped us categorize the individual motivation and barrier variables into broader latent variables (LV) as shown in Table 6.

The EFA results revealed a five-factor structure that characterizes the motivations and barriers related to counterspeech. We applied a robust cut-off threshold of 0.55 for the factor loadings, following [87], and included variable items with strong correlations to the latent variables to ensure that our model was parsimonious and reliable [87], eliminating B9 and M8. The CFA results indicated a good model fit: CFI = 0.952, RMSEA = 0.066 (90% CI = [0.057, 0.076]).

This process resulted in five latent variables associated with counterspeech engagement: **Fear-Driven Inhibition (LV1)**, **Time Concern (LV2)**, **Emotional and Skill Barriers (LV3)**, **Engagement Hesitation (LV4)**, and **Motivation (LV5)** – for a description of these latent variables, see section 6.3 in Appendix.

Table 6: Factor Loadings for Barriers and Motivations Associated with Writing Counterspeech on Social Media

| | Motivation and Barrier Items | LV1 Fear-Driven Inhibition | LV2 Time Concern | LV3 Emotional & Skill Barrier | LV4 Engagement Hesitation | LV5 Motivation |
|---|---|---|---|---|---|---|
| Barriers | B1: Fear of public exposure | **0.694** | | | | |
| | B2: Fear of perpetrator retaliation | **0.882** | | | | |
| | B3: Fear of third-party harassment | **0.848** | | | | |
| | B4: Time Concern | | **1.000** | | | |
| | B5: Emotional burden | | | **0.653** | | |
| | B6: Skill gap | | | **0.565** | | |
| | B7: Engagement unqualified | | | | **0.641** | |
| | B8: Engagement reluctance | | | | **0.593** | |
| | B9: Engagement ineffective | | | | 0.538 | |
| Motivations | M1: Supporting kin | | | | | **0.729** |
| | M2: Supporting others | | | | | **0.842** |
| | M3: Supporting self | | | | | **0.673** |
| | M4: Confronting hate | | | | | **0.815** |
| | M5: Educating ignorance | | | | | **0.793** |
| | M6: Signaling inclusion | | | | | **0.716** |
| | M7: Issue focus | | | | | **0.822** |
| | M8: Venting emotions | | | | | 0.372 |

### 4.2.1 Demographic Characteristics Significantly Influence Motivations and Barriers for Engaging in Online Counterspeech

Our SEM analyses examining the effects of key demographic factors on each latent variable (LV1 – LV5) underlying counterspeech motivations and barriers indicated a strong fit. The chi-square test statistic was 501.092 with 189 degrees of freedom and p < .001. The CFI was 0.915, indicating a good fit. The RMSEA was 0.060, which was within the acceptable range of 0.05 to 0.08.

**Age:** Younger participants experienced higher fear-driven inhibition ($b$ = -.162, $p$ = .001), suggesting that they were more deterred from engaging in counterspeech on social media due to fears of public exposure, perpetrator retaliation, and third-party harassment, compared to their older counterparts. **Gender:** Similarly, women, compared to men also reported higher fear-driven inhibition related to potential retaliation from perpetrators, public exposure, and third-party harassment ($b$ = .228, $p$ < .001). Furthermore, women also reported higher emotional and skill-related barriers - namely the emotional toll of engaging in online counterspeech and uncertainties on how to construct effective counter responses ($b$ = 0.298, $p$ < .001). **Education:** Similarly, participants with higher education levels also reported higher levels of fear-driven inhibition ($b$ = .194, $p$ < .000) and higher emotional burden & skill barriers ($b$ = .181, $p$ = .003) than participants with lower education backgrounds. **Political views:** Participants who were more politically liberal were more motivated to write online counterspeech ($b$ = .150, $p$ = 0.001) and reported less engagement hesitation ($b$ = -.153, $p$ = 0.010) compared to their conservative counterparts.

### 4.3 How do people's motivations and barriers in online counterspeech engagement influence: (a) how satisfied they are with their counterspeech, (b) how difficult they find it to write counterspeech, and (c) how they perceive the effectiveness of their counterspeech (RQ3)?

Figure 2 illustrates the results of our second SEM analysis evaluating the impact of the five latent variables (LV1-LV5) on the three central dependent outcomes: (a) participants' perceived **satisfaction** with their self-authored counterspeech, (b) perceived **difficulty** in writing the counterspeech, and (c) perceived **effectiveness** of their own counterspeech in mitigating the hate speech they were responding to.

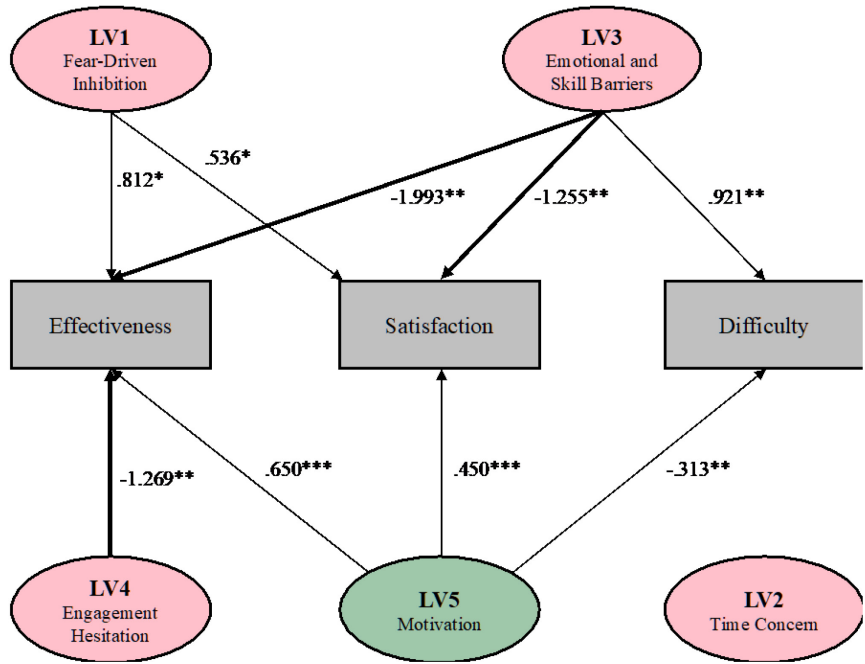

Figure 2: Standardized path coefficients of the structural equation model. We show how the barrier latent variables in red and the motivation latent variable in green affect peoples' perceived satisfaction, difficulty, and effectiveness of their counterspeech. *(\* p < .05; \*\* p < .01; \*\*\* p < .001)*.

The resulting fit indices for our SEM model indicated a good fit. The chi-square statistic was 693.423 ($p < .001$, $df$ = 264). The CFI was 0.900. The RMSEA was 0.060 (90% CI = [0.054, 0.065]), which was within the acceptable range of 0.05 to 0.08. The SRMR was 0.051, which was below the cut-off value of 0.08. In addition to controlling for the control variables in Table 3, we also controlled the model for how hateful the participants perceived the hate speech they were responding to. Below we discuss our results.

**Fear-Driven Inhibition (LV1)**: People who have higher levels of fear of writing counterspeech due to perpetrator retaliation, public exposure, and third-party harassment were significantly more satisfied ($b$ = .536, $p$ = .022) with their self-written counterspeech and perceived their counterspeech to be more effective ($b$ = .812, $p$ = .019) than those who scored lower on the fear-driven inhibition latent variable. One plausible explanation for this is that, due to heightened concerns about negative consequences, users may invest additional effort in crafting their counterspeech [140]. This extra diligence could translate into increased satisfaction and a stronger belief in the effectiveness of their counterspeech

[7].

**Time Concern (LV2)**: Time Concern did not yield statistically significant effects on the outcome variables, suggesting that concerns about the time required for writing counterspeech did not notably influence participants' satisfaction, difficulty, or perceived effectiveness in this context.

**Emotional and Skill Barriers (LV3)**: Those who scored higher on the Emotional and Skill Barriers latent factor were significantly less satisfied with their own counter speech ($b$ = -1.255, $p$ = .002), and experienced more difficulty in writing counterspeech ($b$ = .921, $p$ = .007) in response to the hateful speech they were shown in the survey. These individuals were also significantly less likely to perceive their self-written counterspeech as effective in deterring the hate speech they were responding to ($b$ = -1.993, $p$ = .001).

**Engagement Hesitation (LV4)**: Surprisingly, individuals who scored higher on the Engagement Hesitation latent factor were significantly more likely to perceive their own counterspeech as effective ($b$ = 1.269, $p$ = .008). A plausible explanation for this seemingly paradoxical result could be that those who are hesitant to engage in counterspeech due to feeling unqualified to engage, or are reluctant to converse on social media in general, may set a higher threshold for action. In other words, they may only choose to engage when they believe they have something truly impactful to say. As a result, when they do overcome their hesitation and contribute counterspeech, it may be more thoughtfully crafted, and, thus, such individuals may perceive their counterspeech as more effective.

**Motivation (LV5)**: Participants with stronger motivations for crafting counterspeech not only felt more satisfied with their counter responses ($b$ = .450, $p$ < .001), but also found the writing process less challenging ($b$ = -.313, $p$ = .001). Additionally, they were more confident in the effectiveness of the counterspeech they wrote in response to the hate speech shown in the survey ($b$ = .650, $p$ < .001).

### 4.4 How are people's motivations and barriers in online counterspeech engagement associated with their willingness to use AI assistance (RQ4)?

Results from our first linear regression for RQ4 demonstrate that people who have used ChatGPT (n = 162) in the past are significantly more willing to use AI assistance in writing online counterspeech, compared to those who have never used such tools before (n = 296) ($b$ = .472, $p$ < .001). For specific details, see Table 2A, Appendix. Hence, similar to RQ1, we conducted a subgroup analysis between the two groups. Results are shown in Table 7.

**People with prior experience using AI tools like ChatGPT**: For those who have used ChatGPT in the past, feeling less capable of writing effective counterspeech is a significant motivator for using AI assistance to do so ($b$ = .242, $p$ < .001). Interestingly, this group is also more willing to use ChatGPT when writing counterspeech to signal inclusion to others ($b$ = .145, $p$ = .026), but less inclined to use it when engaging in online counterspeech to stand-up for themselves ($b$ = -.144, $p$ = .039).

**People who have never used AI tools like ChatGPT**: By contrast, among those who have never used ChatGPT, a stronger fear of perpetrator retaliation makes them significantly more willing to use AI for help when writing counterspeech on social media ($b$ = .250, $p$ = .031). However, non-users are less willing to rely on AI assistance when the purpose of engaging in counterspeech is to defend close kin – family and close friends ($b$ = -.265, $p$ = .014). Though feeling incapable of writing effective counterspeech is a significant driver for non-users to rely on AI assistance ($b$ = .214, $p$ = .021), this effect is weaker on non-users compared to those who have used ChatGPT in the past ($b$ = .242, $p$ < .001).

Both prior users and non-users are less willing to use AI for counterspeech writing if they have been targets of online hate speech in the past (prior users: $b$ = -.295, $p$ = .027; non-users: $b$ = -.426, $p$ = .034). Furthermore, political view shows an influence only among prior users, where more liberal views are associated with decreased intention to use AI assistance in crafting counterspeech ($b$ = -.168, $p$ = .004).

Table 7: Results of Subgroup Analyses for Willingness to Use AI Assistance for Writing Online Counterspeech

|  |  | Non Users of ChatGPT (n=162) | | | Prior Users of ChatGPT (n=296) | | |
|---|---|---|---|---|---|---|---|
|  |  | *B* | β | t value | *B* | β | t value |
| Motivations | **M1: Supporting kin** | **-0.265** | **-0.324** | **-2.488*** | 0.085 | 0.093 | 1.134 |
|  | M2: Supporting others | -0.061 | -0.074 | -0.556 | 0.092 | 0.103 | 1.010 |
|  | **M3: Supporting self** | 0.078 | 0.1 | 0.882 | **-0.144** | **-0.162** | **-2.076*** |
|  | M4: Confronting hate | 0.150 | 0.19 | 1.368 | -0.095 | -0.106 | -1.180 |
|  | M5: Educating ignorance | -0.057 | -0.075 | -0.562 | -0.002 | -0.002 | -0.020 |
|  | **M6: Signaling inclusion** | -0.072 | -0.094 | -0.772 | **0.145** | **0.178** | **2.238*** |
|  | M7: Issue focus | 0.171 | 0.201 | 1.497 | 0.001 | 0.001 | 0.017 |
|  | M8: Venting emotions | 0.132 | 0.143 | 1.388 | -0.053 | -0.056 | -0.944 |
| Barriers | B1: Fear of public exposure | -0.045 | -0.058 | -0.450 | 0.087 | 0.097 | 1.366 |
|  | **B2: Fear of perpetrator retaliation** | **0.250** | **0.315** | **2.179*** | 0.122 | 0.132 | 1.550 |
|  | B3: Fear of third-party harassment | -0.069 | -0.088 | -0.604 | -0.058 | -0.068 | -0.808 |
|  | B4: Time Concern | 0.017 | 0.023 | 0.200 | 0.034 | 0.04 | 0.614 |
|  | B5: Emotional burden | 0.070 | 0.087 | 0.776 | -0.064 | -0.071 | -1.024 |
|  | **B6: Skill gap** | **0.214** | **0.261** | **2.337*** | **0.242** | **0.237** | **4.018*** |
|  | B7: Engagement unqualified | -0.114 | -0.137 | -1.200 | 0.026 | 0.028 | 0.400 |
|  | B8: Engagement reluctance | 0.024 | 0.033 | 0.298 | -0.024 | -0.028 | -0.443 |
|  | B9: Engagement ineffective | -0.034 | -0.046 | -0.437 | -0.015 | -0.018 | -0.279 |
| SNS & ChatGPT | **Past experience of online hate speech target** | **-0.426** | **-0.19** | **-2.147*** | **-0.295** | **-0.125** | **-2.222*** |
|  | Frequency of encountering online hate speech | -0.042 | -0.037 | -0.431 | -0.054 | -0.047 | -0.817 |
|  | Social media commenting frequency | 0.060 | 0.057 | 0.627 | 0.132 | 0.112 | 1.857 |
|  | Use of real name on social media | -0.016 | -0.026 | -0.299 | 0.008 | 0.011 | 0.208 |
|  | **Perceived usefulness of ChatGPT** | / | / | / | **0.331** | **0.307** | **5.502*** |
| Demographic | Age | 0.003 | 0.039 | 0.438 | 0.006 | 0.071 | 1.262 |
|  | Gender | 0.022 | 0.01 | 0.115 | -0.005 | -0.002 | -0.040 |
|  | Ethnicity | 0.176 | 0.082 | 0.950 | 0.109 | 0.047 | 0.842 |
|  | Education level | -0.038 | -0.026 | -0.299 | 0.092 | 0.057 | 1.034 |
|  | Sexual orientation | -0.416 | -0.147 | -1.534 | 0.115 | 0.042 | 0.724 |
|  | **Political views** | 0.086 | 0.092 | 0.957 | **-0.168** | **-0.169** | **-2.875**** |
|  | (Intercept) | 1.477 | / | 1.904 | 0.318 | / | 0.533 |
| R-squared |  | 0.2132 | | | 0.3157 | | |
| F-test |  | F(27, 137) = 1.345, p = .138 | | | F(28, 267) = 4.399, p < .001 | | |

### 4.4.1 Qualitative Analysis: Motivations and Reservations for Using AI for Counterspeech Writing

Our qualitative analysis of participants' open responses resulted in a total of six themes associated with why people would or would not use AI assistance for writing online counterspeech. Tables 8 and 9 show the main themes that emerged, along with illustrative examples and the proportion of responses that fell into each theme. We had three raters who independently coded the participants' open responses into the themes that were identified. The overall Cohen's kappa coefficient for our analysis was 0.854, with a 95% confidence interval of 0.817 to 0.891. This indicates a very good level of agreement among the raters. Because some user statements contained more than one theme, we coded them into multiple categories. Therefore, the total percentages of the themes exceed 100%. We discuss each theme in detail below.

Table 8: Reasons for Using AI to Write Online Counterspeech (38.5%)

| Themes | Illustrative Quotes | % |
|---|---|---|
| Efficiency and Convenience | • *I think it's better and faster at putting together coherent sentences that get my point across than myself.*<br>• *It can save time and effort compared to writing a response from scratch.*<br>• *It can give me ideas quickly and I can elaborate with my own perspective of the facts.*<br>• *I think it would save me time and energy, maybe it could help me to better learn the skill so I could use it more.* | 24.7% |
| Less Emotional Burden | • *It would take all of the emotional work out of it for me.*<br>• *It saves you the stress and irritation of having to respond to an ignorant person.*<br>• *I feel like ChatGPT would be able to refute it with facts and logic in better ways than me, because I feel like counterspeech is an emotional burden on me and I get overwhelmed.*<br>• *I would probably have the AI help, partially because I just don't have the energy for that sort of thing anymore.* | 11.6% |
| Access to Larger Knowledge Base & Better Articulation | • *It can help me find the right words and vocabulary to express my thoughts more clearly and eloquently.*<br>• *It has access to a huge breadth of knowledge that I don't, so it can provide data and facts to make my argument stronger.*<br>• *ChatGPT would be able to assist me with my argument in order to make my counterspeech more effective.*<br>• *I would use it to help get my statement across in a much clearer way. Also to help me make sure that the information I am writing about is correct.*<br>• *ChatGPT has a broad database full of statistics and information, and I feel as though it would create the most effective counterspeech because of that. It has nearly all of the information in the world within it, it would certainly make an argument more efficient than I probably could.* | 2.2% |

**Efficiency and Convenience**: Participants emphasized how using AI can *"save time and effort compared to writing a response from scratch"*, thereby making the writing process faster for them. Some participants also highlighted that AI could help them more easily come up with ideas that they can elaborate on themselves and provide them with useful strategies for writing effective counterspeech that they can use later on.

**Less Emotional Burden:** Many participants highlight that AI tools like ChatGPT could help alleviate many of the negative emotions, such as anger and frustration, that arise when writing counterspeech. For example, one participant notes: *"it can save the stress and irritation of responding to an ignorant person"*.

**Access to Larger Knowledge Base & Better Articulation:** Many participants also underscored the ability of AI to not only help them express themselves more clearly, but also quickly provide supporting evidence for arguments. For example, participants state that AI tools like ChatGPT can help them *"find the right words and vocabulary to express [their] thoughts more clearly and eloquently,"* as well as *"provide data and facts to make [their] argument stronger"*.

**Authenticity and Ethical Concerns**: The most common reservations for using AI assistance in writing counterspeech are authenticity and ethical concerns. Some voiced feelings of cheating and unease, with one participant stating that *"Using ChatGPT to make counterspeech and then posting it as if it were [their] own is lying and unethical at *best.*". Moreover, others expressed worries that AI usage detaches their words from personal ownership, with a participant saying that they would want their counterspeech "*to be in their own words and thoughts*".

**Lack of Emotional, Human, or Personal Touch**: Many participants raised doubt about AI's ability to mimic human emotions such as empathy, with a participant saying that they believe AI "*can use logic but not empathy to write counterspeech*". Additionally, participants also mentioned AI's lack of ability to capture their personal experiences or "*fully express what [they] want to express*".

**Lack of Familiarity or Trust in AI**: Many participants also seem to have a general distrust of AI technology, with one participant stating they "*don't think ChatGPT and AI in general is quite the 'do it all' answer everyone acts like it is*". Others cite their lack of familiarity with AI tools as the primary reason for their distrust. Moreover, many also recognize that AI may not be "*100% accurate or correct*".

Table 9: Reservations Against Using AI to Write Online Counterspeech (71.7%)

| Themes | Illustrative Quotes | % |
|---|---|---|
| Authenticity and Ethical Concerns | • *If I were being graded for a counterspeech, it would be cheating to have anyone or anything write it for me.*<br>• *It's not my voice. It's not my perspective. Personally I'd be ashamed to utilize Artificial Intelligence for counterspeech.*<br>• *Because then it wouldn't even be MY counterspeech. Why would I use AI to write MY opinion? It's stupid.*<br>• *If my statement were to be judged by others, I would want the statement to be in my own words using my own thoughts.*<br>• *Using ChatGPT to make counterspeech and then posting it as if it were my own is lying and unethical at best.*<br>• *I would want to build the skills to effectively and reliably write such speech myself.* | 33.0% |
| Lack of Emotional, Human, or Personal Touch | • *This needs human sentiment with human feelings behind them. ChatGPT AI may get there but it's not there yet.*<br>• *I think I could write it better because I can use personal experiences and my emotions to hopefully make the perpetrator really think about it.*<br>• *It can use logic, but not empathy, to write counterspeech.*<br>• *I would rather tailor my response to be exactly what I'm thinking. I'm not sure it could fully express what I want to express, and it may lack nuance.*<br>• *It doesn't come from the heart.* | 26.0% |
| Lack of Familiarity or Trust in AI | • *I'm not familiar with it, hence my trust level in its performance is low.*<br>• *I don't think ChatGPT has enough understanding of how internet commenting dynamics work.*<br>• *I don't think ChatGPT and AI in general is quite the "do it all" answer everyone acts like it is.*<br>• *It's not always 100% accurate or correct and could cause issues if you post it as counterspeech and it turns out to be incorrect.*<br>• *I trust my own words more than a robot's.* | 12.7% |

## 5 DISCUSSION

### 5.1 Motivations and Barriers in Online Counterspeech

#### 5.1.1 Main Deterrent of Online Counterspeech Engagement: General Reluctance to Engage on Social Media

RQ1 results show peoples' general reluctance to engage on social media tends to outweigh other motivations for engaging in online counterspeech. This aligns with prior research in HCI that identifies reluctance towards social media interaction as a contributing factor to online bystander effect [39, 138, 147]. For this reason, while a majority of users observe online harassment, less than a third choose to intervene [6]. Nevertheless, our findings go a step further by examining key emotional and psychological factors that influence this behavior.

#### 5.1.2 Main Driver of Online Counterspeech Engagement: Prior Victimization

Studies in online bystander intervention show that prior victimization is a key predictor of bystander action [129]. RQ1 results confirm these insights, demonstrating that having been a target of online hate speech in the past is the strongest predictor that drives people to frequently engage in online counterspeech. We provide further nuance to prior scholarship by showing how counterspeech motivations and barriers in fact, significantly vary between former targets and non-targets. For targets, the perceived ineffectiveness of their counterspeech is the most significant barrier to engaging in counterspeech, while non-targets are primarily deterred by their reluctance to engage on social media in general. These findings are supported by existing work that suggests victims know well when an intervention may or may not be effective due to the memory of their own experiences as well as bystander effects on social media [70, 129]. With respect to peoples' motivations, Costello et al. (2016) found that past victims of online hate are more than three times as likely to defend fellow victims [34]. Our results provide context to this research, showing that targets engage in online counterspeech more often when they are primarily motivated to confront a hateful person or their behavior, while this motivation is insignificant for non-targets.

#### 5.1.3 Multi-Item Survey Scale for Measuring Online Counterspeech Motivations and Barriers

While prior research in cyberbullying has created numerous scales to measure bystander motivations [139, 147], most are not generalizable to context of online counterspeech due to differences highlighted in prior research [129]. Furthermore, such studies are often based on children and adolescents [9, 10, 114], while our study focuses on adults. To the best of our knowledge, our work is the first to provide a comprehensive set of survey scales for understanding online counterspeech motivation and barriers. Furthermore, as shown in RQ2, each motivation and barrier scale can be constructed into five key latent variables (LV1 – LV5). Researchers use these variable items both individually and as latent factors in future studies.

#### 5.1.4 Demographic Variances in Counterspeech Motivations and Barriers

Using our latent variables (LV1 -LV5), we demonstrate key demographic variances in counterspeech motivations and barriers (RQ2). We found that age was negatively associated with several barriers, such as fear, emotional and skill barriers, and engagement hesitation, meaning that younger adults are more likely to be deterred by these factors. Our findings are consistent with prior HCI research that highlights nuanced differences in how older versus younger adults perceive online risk [56] and safety [2, 3]. Another notable finding in our work is that women not only face a higher fear of retaliation from the perpetrator and third parties, but also fear being publicly exposed when countering hate through online counterspeech. This aligns with prior research documenting how women cope with online harassment by adopting gendered defensive strategies [23, 92]. Particularly, women tend to experience more depression and anxiety after being harassed online, partially explaining their heightened fear of retaliation [23, 108]. Moreover, we found that women reported lower self-efficacy in their counterspeech writing skills and greater emotional burden concerns than men. Recent studies suggests how women who are targeted by online harassment become more cautious in expressing their opinions publicly, as they tend to normalize harassment, self-censor, or withdraw from online spaces to avoid further harm [23]. Our work provides evidence that these factors may deter women from engaging in online counterspeech. In addition, research has shown that more educated individuals may better understand the potential risks and difficulties of public online engagement [11]. However, our findings show that participants with higher education levels report higher barriers stemming from fear-driven inhibition as well as emotional burden and skill barriers. Concerns of potential risks can prevent more educated individuals from writing counterspeech, even though they may have the necessary skills and knowledge to do so.

*5.1.5 Demographic Factors That Shape Counterspeech Writing Experiences*

RQ3 results show that compared to other demographic groups, younger, female, more educated users tend to feel more satisfied towards their self-written counterspeech and perceive their counterspeech to be more effective. Interestingly, this group not only encounters hate speech more frequently, but is also more likely to have been a target of online hate speech in the past. Our results also show a positive correlation between the frequency of online hate speech exposure and fear-driven inhibition, meaning that the more often one encounters online hate speech, the stronger their fear-driven inhibitions to engage in counterspeech. As previously discussed, more exposure to online hate may increase peoples' awareness of the potential threats and challenges of countering it [11, 23, 94]. This may allow individuals to leverage their personal experiences and knowledge [62, 142], leading them to write more satisfying and effective counterspeech. Another explanation is that greater awareness of the potential risks and difficulties in countering online hate may motivate users to put more effort and care into crafting their responses [140]. This extra diligence could lead to more satisfaction and a stronger belief in the impact of their counterspeech [7]. Further research could explore these relationships in more depth.

*5.1.6 The Role of AI Assistance in Counterspeech Writing: AI-mediated Counterspeech*

AI-mediated communication can be defined as interpersonal communication that is not simply transmitted by technology but augmented or even generated by algorithms to achieve specific communicative or relational outcomes [75]. Since trust is fundamental to human relationships and manifests in collaborative behaviors such as a willingness to depend on and share information, previous research has placed a significant emphasis on examining people's perceptions of trustworthiness of messages generated with AI assistance [75, 93]. Our qualitative findings provide further nuance to prior research by revealing a tension between individuals' hesitations and motivations for using AI assistance in crafting online counterspeech. While the primary motivation for adopting such technology is rooted in its utilitarian advantages, reservations predominantly revolve around issues of trust. Most participants expressed distrust in AI's ability to accurately convey their emotions as well as the credibility of the information it presents. Given that prior research has found that peoples' trust in AI generated text decreases as the level of AI agency (the degree of autonomous content generation and decision-making by AI) increases, we propose AI-*mediated* counterspeech as an alternative to AI-*generated* counterspeech [93, 102]. We discuss specific design implications in 5.2.3.

### 5.2 Design Implications

#### 5.2.1 Incentivizing Participation Through Recognition

Understanding the barriers that deter users from engaging in online counterspeech is a crucial step for developing effective design strategies to encourage proactive participation and intervention in online spaces. People's reluctance to engage on social media is one of the strongest barriers against participating in online counterspeech (RQ1). Prior research has shown that incentivizing user contribution through community acknowledgment through badge systems can increase engagement among those who prefer to be lurkers [21]. For instance, digital badges can function as an award mechanism through visible symbols of achievement and recognition [46, 91]. StackOverflow [46] and Reddit [91] use a system where users earn badges for varying levels of contribution, such as providing helpful answers or engaging in community moderation. Researchers have shown that these badges not only increase engagement, but also a sense of responsibility and motivation to ensure adherence to community norms in an empathetic manner [21]. Relatedly, our work shows that confronting hate and supporting others are important motivators for engaging in counterspeech (RQ1). Building on this, online platforms could introduce badges for users who engage in constructive counterspeech. Visible badges on user profiles could both recognize individual contributions to counterspeech efforts and potentially encourage broader participation within the community to engage in constructive online counterspeech.

#### 5.2.2 Mitigating Fear Through Personalized Interactions with Others' Responses to User's Counterspeech

A notable finding in our study is that younger individuals, females, and those who are more educated feel greater fear related to engaging in online counterspeech, and that this fear is correlated with the amount of hateful content they encounter online (RQ2). To address this, HCI researchers can design personalized features to empower users to manage how they interact with responses, particularly harmful and retaliatory ones, to their counterspeech. Supporting this approach, prior HCI studies have identified a clear preference for personalized content moderation for harmful content, especially among users who have been victims of online hate [31, 123]. For instance, transgender Twitter users, who regularly encounter transphobic content online, appreciate the ability to automatically filter out posts containing offensive words specific to their personal preferences, eliminating the need to repetitively mute offenders or posts [71, 77]. In the context of reducing fear towards hateful responses to one's counterspeech, similar design features could be implemented to provide users with a more personalized approach in their ability to moderate responses to their counterspeech, while minimizing exposure to harmful responses from retaliators or third-parties. For example, users could have the choice to either subject these responses to pre-publication moderation or to engage an automatic filtering mechanism based on personally curated keywords for explicitly offensive reactions. Integrating personalized keyword filters or those based on community norms to blur out aggressive or offensive responses to one's counterspeech could potentially lessen the impact and, consequently, the fear of retaliation, thereby creating a safer and more controlled environment for users to engage in counterspeech.

#### 5.2.3 Towards Better Understanding of Human-AI Collaboration in Co-Writing Online Counterspeech

Research shows that when users work together with AI toward a common goal, they may treat AI as a collaborative partner instead of a tool [59, 154]. Similarly, our qualitative analysis in RQ4 show that participants' willingness to use AI for counterspeech writing was often based on expectations of the AI's role and the degree of AI involvement in the writing process. For example, many expressed a preference to use AI to help them brainstorm ideas rather than having AI completely write the counterspeech for them. For such users, LLM-powered AI systems may be designed to facilitate brainstorming sessions, by allowing users to input words or phrases as fragments of their thoughts, or details they wish to provide based on personalized experiences. In response, the system may provide feedback and suggestions based on its training and understanding of what is considered constructive counterspeech. Furthermore, participants often indicated overcoming emotional burden or making sure than they do not sound overly angry in their responses as primary reasons for using AI assistance in counterspeech writing. However, despite wanting to use AI for better articulation of their emotions and thoughts, users also expressed concerns around conveying authenticity. Researchers in the HCI community, particularly those focusing on AI-assisted co-writing [55, 89], could examine these issues in future work by exploring ways to design human-AI interactions that can help users convey tone and emotion in their counterspeech, without diminishing the user's sense of personal agency and authenticity in their counterspeech writing process.

## 6 LIMITATIONS

While we have randomly assigned hate posts across diverse topics (gender, religion, disability, sexual orientation, and race) among our survey participants, we understand that such topics may impact each participant and their responses differently. We recognize this as a limitation of our study, and plan to conduct a more detailed investigation in our future work to understand how these different topics in hate posts affect peoples' motivations and barriers when it comes to engaging in counterspeech to these posts, as well as their actual counter-responses to topically diverse hate posts.

We understand that differences in participants' primary social media platform of choice may influence their survey responses. Therefore, we included this variable as a control in the linear regressions of RQ1 and RQ4 to verify the impact. However, our analysis revealed that this variable neither altered the significance nor the direction of the regression coefficient; hence, we moved these results to Table A3 (Appendix) and removed this non-significant variable to simplify the regression models in the main text. Finally, participants were asked to write a counterspeech in response to real-life posts containing hate speech in a survey setting, which is different from responding to a hateful post in real life. Hence, this may impact how they write counterspeech in addition to how they evaluate it.

## 7 CONCLUSION

Our work investigates the various motivations and barriers that underly online counterspeech engagement. To this end, we developed and validated a multi-item scale to assess these factors, demonstrating its significant influence on both the experience of writing counterspeech and people's perceptions towards their self-authored counterspeech. These measures can be used both as individual and latent factors, providing a scale that can be operationalized in future studies in relevant areas of scholarship. Using these latent variables, we demonstrate key demographic variances in counterspeech motivations and barriers, which have not been studied in prior research. Furthermore, we contribute to the emerging understanding of factors influencing people's openness to using AI assistance for crafting online counterspeech: while AI can assist in crafting responses, it cannot replace the human element essential for genuinely empathetic and contextually appropriate counterspeech. This finding underscores the necessity of human-AI collaboration, where AI's limitations are complemented by human insight and judgment. Our findings can guide tech firms and researchers in better understanding the role of AI in helping users combat hate speech on online platforms. This is especially timely as the implementation of AI technologies in facilitating online discourse is becoming more prominent.

# 8 APPENDICES

## 8.1 Survey Questions

This section presents the key questions from the survey; for the complete list of questions, please refer to the link available on the Open Science Framework (OSF) [3].

**Screen question:** Imagine you are a user of an online group on social media. Another user (perpetrator) in the group posted the following: *[hateful post]*. Do you consider this post to be hateful? (Yes, No)

**Satisfaction:** How satisfied are you with the counterspeech that you've written? (Extremely dissatisfied, Somewhat dissatisfied, Neither satisfied nor dissatisfied, Somewhat satisfied, Extremely satisfied)

**Difficulty:** How difficult was it to write this counterspeech? (Extremely difficult, Somewhat difficult, Neither easy nor difficult, Somewhat easy, Extremely easy)

**Self-perceived effectiveness:** How effective do you think your counterspeech would be in preventing the perpetrator from engaging in further hateful behavior? (Not effective at all, Slightly effective, Moderately effective, Very effective, Extremely effective)

**Prior experience of online hate speech target:** Have you been a target of hateful speech on the internet? (Yes, No)

**Barriers:** How much do the following factors prevent you from writing a counterspeech on social media? (None at all, A little, A moderate amount, A lot, A great deal)

- B1: I fear being publicly exposed.
- B2: I'm afraid of retaliation from the perpetrator.
- B3: I'm afraid that I will be harassed by people (other than the perpetrator).
- B4: I don't want to spend time on this.
- B5: Writing a counterspeech is emotionally burdensome.
- B6: I don't know how to write an effective counterspeech.
- B7: I feel that it's not my place to engage in counterspeech.
- B8: I don't like to engage in social media conversations.
- B9: I feel that my counterspeech would not make a difference.

**Motivations:** How much do the following factors motivate you to write a counterspeech on social media? (None at all, A little, A moderate amount, A lot, A great deal)

- M1: When I feel the need to stand up for people I care about (e.g., family, close friends).
- M2: When I feel the need to stand up for people in general.
- M3: When I feel the need to stand up for myself.
- M4: When I want to confront a hateful person or behavior.
- M5: When I want to educate an ignorant person.
- M6: To signal that I stand for inclusion.
- M7: When it concerns issues or topics I care about.
- M8: When I want to blow off steam.

**Frequency of writing online counterspeech**: How often do you write counterspeech on social media? (Never, Rarely, Sometimes, Often, Frequently)

---

[3] https://osf.io/rzmg3/?view_only=6b2fd3a3d42b4b25a37f014612fac18a

**Prior use of ChatGPT:** Have you used ChatGPT before? ChatGPT is an artificial intelligence tool that allows you to have human-like conversations by generating human-like responses to text-based inputs. ChatGPT can answer questions, and assist you with tasks such as composing emails, essays, and code. (Yes, No)

**Perceived usefulness of ChatGPT:** How useful do you find ChatGPT? (Not at all useful, Slightly useful, Moderately useful, Very useful, Extremely useful)

**Willingness to use ChatGPT to help write counterspeech:** If you were writing a counterspeech on social media, would you use artificial intelligence technology like ChatGPT to help you write it?

**Social media platforms:** What social media platform do you use most often? (Facebook, Instagram, Twitter/X, Linkedin, Reddit, YouTube, TikTok, Snapchat, Other, I don't currently use social media)

**Social media usage:** I have used social media to. (Stay informed on current events/news, Shop, Learn, Socialize, Entertain myself, Advocate for social issues I care about)

**Post frequency:** How often did you post on social media? (Never, Rarely, Sometimes, Often, Always)

**Comment frequency:** How often did you comment on content that you encountered on social media? (Never, Rarely, Sometimes, Often, Always)

**Online Anonymous:** Did you use your real name on social media? (Never, Rarely, Sometimes, Often, Always)

**Social media experience:** I encountered the following on social media. (Never, Rarely, Sometimes, Often, Always)
- Content that I disagree with.
- Content that I find hateful.
- Content that I find controversial.

### 8.2 Description of Latent Variables

**Fear-Driven Inhibition (LV1):** This latent factor captures a range of fears related to engaging in online counterspeech. It includes concerns about public exposure (B1), fears of retaliation from the perpetrator (B2), and fear of harassment from third parties (B3).

**Time Concern (LV2):** This one-item latent variable is the barrier associated with time concern in engaging in online counterspeech.

**Emotional and Skill Barriers (LV3):** This dual-variable factor addresses both the emotional toll and skill-related concerns when engaging in online counterspeech. The factor consists of the emotional burden of engaging in counterspeech (B5) along with uncertainty regarding how to write an effective counterspeech (B6).

**Engagement Hesitation (LV4):** This latent variable captures engagement-related barriers in engaging in online counterspeech, which includes feeling unqualified to engage in counterspeech (B7), and a general reluctance to engage in social media conversations as a reason for not engaging in online counterspeech (B8).

**Motivation (LV5):** Except for venting emotions (M8), this latent variable embodies all motivation variables for engaging in counterspeech – including standing up for kin (M1), others in general (M2), and oneself (M3), confronting hate (M4), educating the ignorant (M5), signaling inclusion (M6), focusing on issues of personal importance (M7).

### 8.3 Influence of Social Media Platforms

We also considered the influence of different social media platforms on counterspeech. We asked participants to indicate which platforms they currently mainly use from a list of 10 options: Facebook, Instagram, Twitter (X), Linkedin, Reddit, YouTube, TikTok, Snapchat, Other, and I don't currently use social media. We used Facebook as the reference level. The results are shown in Table A3. However, we found that these differences were not substantial enough to significantly affect the overall findings. Therefore, we did not include the platform variable in our main analysis.

## 8.4 Tables

Table A1. Participant Demographics (N = 458)

| Factor | Category | N |
|---|---|---|
| Age group | 18-30 | 138 |
|  | 31-60 | 293 |
|  | 61-81 | 27 |
| Gender | Male | 226 |
|  | Female | 232 |
|  | Other or prefer not to answer | 0 |
| Ethnicity | Majority | 234 |
|  |     White | 234 |
|  | Minority | 224 |
|  |     Asian | 55 |
|  |     Black | 110 |
|  |     Hispanic | 53 |
|  |     Other | 6 |
| Education level | Less than high school or high school graduate | 65 |
|  | Some college or 2-year degree | 154 |
|  | 4-year degree or higher | 239 |
| Sexual orientation | Heterosexual | 359 |
|  | Non-Heterosexual | 99 |
| Political views | Very Conservative | 28 |
|  | Conservative | 91 |
|  | Moderate | 121 |
|  | Liberal | 134 |
|  | Very Liberal | 84 |

Table A2. Linear Regression Results of Willingness to Use ChatGPT to Write Counterspeech on Social Media

| | | *B* | β | Std. Error | t value | VIF |
|---|---|---|---|---|---|---|
| Motivations | M1: Supporting kin | -0.041 | -0.046 | 0.061 | -0.673 | 2.482 |
| | M2: Supporting others | 0.023 | 0.027 | 0.070 | 0.336 | 3.292 |
| | M3: Supporting self | -0.014 | -0.016 | 0.054 | -0.258 | 2.085 |
| | M4: Confronting hate | -0.010 | -0.012 | 0.065 | -0.160 | 2.953 |
| | M5: Educating ignorance | -0.017 | -0.021 | 0.061 | -0.277 | 2.870 |
| | M6: Signaling inclusion | 0.102 | 0.127 | 0.053 | 1.948 | 2.221 |
| | M7: Issue focus | 0.083 | 0.09 | 0.067 | 1.234 | 2.783 |
| | M8: Venting emotions | -0.045 | -0.048 | 0.049 | -0.930 | 1.382 |
| Barriers | B1: Fear of public exposure | 0.071 | 0.082 | 0.054 | 1.320 | 2.012 |
| | **B2: Fear of perpetrator retaliation** | **0.187** | **0.211** | **0.065** | **2.891**** | **2.788** |
| | B3: Fear of third-party harassment | -0.104 | -0.124 | 0.061 | -1.711 | 2.767 |
| | B4: Time Concern | 0.021 | 0.026 | 0.046 | 0.458 | 1.705 |
| | B5: Emotional burden | -0.076 | -0.087 | 0.050 | -1.515 | 1.722 |
| | **B6: Skill gap** | **0.192** | **0.203** | **0.050** | **3.813**** | **1.484** |
| | B7: Engagement unqualified | 0.024 | 0.026 | 0.054 | 0.449 | 1.803 |
| | B8: Engagement reluctance | 0.002 | 0.003 | 0.045 | 0.055 | 1.667 |
| | B9: Engagement ineffective | -0.012 | -0.015 | 0.045 | -0.273 | 1.513 |
| SNS & ChatGPT | **Prior use of ChatGPT** | **0.472** | **0.197** | **0.109** | **4.334**** | **1.089** |
| | **Past experience of online hate speech target** | **-0.375** | **-0.16** | **0.109** | **-3.427**** | **1.150** |
| | Frequency of encountering online hate speech | -0.036 | -0.031 | 0.055 | -0.650 | 1.168 |
| | Social media commenting frequency | 0.087 | 0.076 | 0.057 | 1.515 | 1.320 |
| | Use of real name on social media | 0.007 | 0.011 | 0.031 | 0.237 | 1.138 |
| Demographic | Age | 0.002 | 0.019 | 0.004 | 0.402 | 1.214 |
| | Gender | 0.020 | 0.009 | 0.110 | 0.185 | 1.219 |
| | Ethnicity | 0.161 | 0.07 | 0.108 | 1.489 | 1.164 |
| | Education level | 0.055 | 0.034 | 0.075 | 0.731 | 1.166 |
| | Sexual orientation | 0.091 | 0.033 | 0.138 | 0.659 | 1.292 |
| | **Political views** | **-0.118** | **-0.12** | **0.049** | **-2.403*** | **1.308** |
| | (Intercept) | 1.029 | / | 0.470 | 2.187* | / |

Adjusted R-squared = 0.4532; $F(27, 430) = 15.03$, $p < .001$

Table A3. Effects of Social Media Platforms on All Dependent Variables

|  | Counterspeech Writing Frequency | | Willingness to Use AI Assistance | |
| --- | --- | --- | --- | --- |
|  | B | P | B | P |
| Facebook (Ref level) | / | / | / | / |
| Instagram | 0.148 | 0.229 | -0.150 | 0.395 |
| Twitter (X) | 0.085 | 0.554 | -0.059 | 0.774 |
| Linkedin | -0.109 | 0.810 | -0.221 | 0.732 |
| Reddit | 0.124 | 0.432 | -0.183 | 0.422 |
| YouTube | 0.045 | 0.726 | -0.094 | 0.608 |
| TikTok | -0.131 | 0.369 | 0.127 | 0.542 |
| Snapchat | 0.144 | 0.659 | -0.445 | 0.338 |
| Other | -0.430 | 0.248 | -1.151 | 0.031 |
| I don't currently use social media | -0.669 | 0.391 | -0.011 | 0.992 |
| (Intercept) | 1.027 | / | 1.135 | / |

Table A4. Relationship Between Latent Variables and Covariates in the Structural Equation Model

| IVs | Covariates | Estimate | Std. Est | Std. Error | z value | P-value |
|---|---|---|---|---|---|---|
| Fear-Driven Inhibition | Age | -1.948 | -0.162 | 0.616 | -3.160 | 0.001* |
|  | Gender | 0.114 | 0.228 | 0.024 | 4.723 | 0.000* |
|  | Ethnicity | -0.019 | -0.040 | 0.022 | -0.853 | 0.394 |
|  | Education level | 0.136 | 0.194 | 0.034 | 3.994 | 0.000* |
|  | Sexual orientation | 0.025 | 0.064 | 0.019 | 1.359 | 0.174 |
|  | Political views | 0.009 | 0.008 | 0.052 | 0.177 | 0.860 |
|  | Past experience of online hate speech target | 0.016 | 0.035 | 0.022 | 0.742 | 0.458 |
|  | Frequency of encountering hateful content | 0.117 | 0.126 | 0.045 | 2.625 | 0.009* |
|  | Social media commenting frequency | 0.014 | 0.015 | 0.045 | 0.320 | 0.749 |
|  | Use of real name on social media | -0.055 | -0.034 | 0.076 | -0.718 | 0.473 |
| Time Concern | Age | 0.086 | 0.005 | 0.859 | 0.101 | 0.920 |
|  | Gender | -0.033 | -0.047 | 0.032 | -1.034 | 0.301 |
|  | Ethnicity | -0.022 | -0.031 | 0.032 | -0.689 | 0.491 |
|  | Education level | 0.082 | 0.081 | 0.047 | 1.762 | 0.078 |
|  | Sexual orientation | 0.013 | 0.023 | 0.027 | 0.506 | 0.613 |
|  | Political views | -0.050 | -0.031 | 0.075 | -0.670 | 0.503 |
|  | Past experience of online hate speech target | 0.015 | 0.021 | 0.032 | 0.466 | 0.641 |
|  | Frequency of encountering hateful content | -0.032 | -0.023 | 0.063 | -0.501 | 0.616 |
|  | Social media commenting frequency | -0.269 | -0.190 | 0.067 | -4.023 | 0.000* |
|  | Use of real name on social media | -0.144 | -0.059 | 0.110 | -1.302 | 0.193 |
| Emotional & Skill Barrier | Age | -1.270 | -0.115 | 0.558 | -2.277 | 0.023* |
|  | Gender | 0.096 | 0.298 | 0.023 | 4.108 | 0.000* |
|  | Ethnicity | -0.034 | -0.083 | 0.021 | -1.616 | 0.106 |
|  | Education level | 0.092 | 0.181 | 0.031 | 2.997 | 0.003* |
|  | Sexual orientation | -0.002 | -0.007 | 0.017 | -0.145 | 0.884 |
|  | Political views | 0.016 | 0.016 | 0.051 | 0.310 | 0.757 |
|  | Past experience of online hate speech target | 0.018 | 0.045 | 0.020 | 0.904 | 0.366 |
|  | Frequency of encountering hateful content | 0.069 | 0.085 | 0.041 | 1.699 | 0.089 |
|  | Social media commenting frequency | -0.116 | -0.139 | 0.044 | -2.608 | 0.009* |
|  | Use of real name on social media | -0.074 | -0.053 | 0.072 | -1.031 | 0.303 |
| Engagement Hesitation | Age | -1.403 | -0.124 | 0.624 | -2.248 | 0.025* |
|  | Gender | 0.025 | 0.059 | 0.023 | 1.063 | 0.288 |
|  | Ethnicity | -0.009 | -0.021 | 0.023 | -0.393 | 0.694 |
|  | Education level | 0.065 | 0.107 | 0.034 | 1.944 | 0.052 |
|  | Sexual orientation | -0.009 | -0.025 | 0.019 | -0.461 | 0.645 |
| | Political views | -0.153 | -0.134 | 0.055 | -2.387 | 0.010* |
| | Past experience of online hate speech target | -0.007 | -0.017 | 0.022 | -0.322 | 0.747 |
| | Frequency of encountering hateful content | 0.022 | 0.027 | 0.045 | 0.500 | 0.617 |
| | Social media commenting frequency | -0.229 | -0.270 | 0.051 | -4.513 | 0.000* |
| | Use of real name on social media | -0.003 | -0.002 | 0.079 | -0.037 | 0.970 |
| Motivation | Age | 0.992 | 0.080 | 0.555 | 1.786 | 0.074 |
| | Gender | 0.016 | 0.035 | 0.021 | 0.793 | 0.428 |
| | Ethnicity | 0.020 | 0.043 | 0.021 | 0.969 | 0.332 |
| | Education level | -0.026 | -0.039 | 0.030 | -0.868 | 0.385 |
| | Sexual orientation | 0.011 | 0.028 | 0.017 | 0.622 | 0.534 |
| | Political views | 0.170 | 0.150 | 0.050 | 3.401 | 0.001* |
| | Past experience of online hate speech target | 0.061 | 0.134 | 0.021 | 2.937 | 0.003* |
| | Frequency of encountering hateful content | 0.163 | 0.181 | 0.042 | 3.859 | 0.000* |
| | Social media commenting frequency | 0.265 | 0.286 | 0.046 | 5.716 | 0.000* |
| | Use of real name on social media | 0.142 | 0.090 | 0.071 | 1.988 | 0.047* |